\documentclass[journal]{IEEEtran}
\IEEEoverridecommandlockouts
\usepackage{cite}
\usepackage{bm}
\usepackage{amsmath,amssymb,amsfonts}
\usepackage{graphicx}
\usepackage{multirow}
\usepackage{bm}
\usepackage{float}
\usepackage{amsmath,amsfonts}
\usepackage{algorithmic}
\usepackage{algorithm}
\usepackage{array}
\usepackage[caption=false,font=normalsize,labelfont=sf,textfont=sf]{subfig}
\usepackage{textcomp}
\usepackage{stfloats}
\usepackage{url}
\usepackage{verbatim}
\usepackage{cite}
\usepackage{amsmath, cite}
\usepackage{graphicx}
\usepackage{enumitem}

\usepackage{array}
\usepackage{textcomp}
\usepackage{epstopdf}
\usepackage{caption}
\usepackage[dvipsnames]{xcolor}
\usepackage{booktabs}
\renewcommand{\baselinestretch}{0.955} 

\usepackage[printonlyused,withpage]{acronym}

\newcommand{\bs}[1]{\ensuremath{\boldsymbol{#1}}}
\newcommand{\mrm}[1]{\ensuremath{\mathrm{#1}}}
\newcommand{\diag}{\ensuremath{\mathrm{diag}}}

\acrodef{BS}{base station}
\acrodef{CSI}{channel state information}
\acrodef{ZF}{zero-forcing}
\acrodef{UE}{user equipment}
\acrodef{EM}{expectation maximization}
\acrodef{UL}{uplink}
\acrodef{DL}{downlink}
\acrodef{TDD}{time-division duplexing}
\acrodef{FDD}{frequency-division duplexing}
\acrodef{ZF}{zero-forzing}
\acrodef{LMMSE}{linear minimum mean squared error}
\acrodef{MRC}{maximum ratio combining}
\acrodef{BBU}{baseband unit}
\acrodef{DPD}{digital pre-distortion}
\acrodef{OTA}{over-the-air}
\acrodef{SNR}{signal to noise ratio}
\acrodef{TX}{transmit}
\acrodef{RX}{receive}
\acrodef{AWGN}{additive white Gaussian noise}
\acrodef{MIMO}{multiple-input multiple-output}
\acrodef{LS}{least-squares}
\acrodef{MVU}{minimum-variance unbiased}
\acrodef{EM}{expectation-maximization}
\acrodef{MSE}{mean square error}
\acrodef{CRLB}{Cramer-Rao Lower Bound}

\begin{document}
\title{Over-the-Air DPD and Reciprocity Calibration in Massive MIMO and Beyond \\
}

\author{\IEEEauthorblockN{Ashkan Sheikhi,~\IEEEmembership{Member,~IEEE,} Ove Edfors,~\IEEEmembership{Senior Member,~IEEE,} and Juan Vidal Alegr\'{i}a,~\IEEEmembership{Member,~IEEE,}.}
\thanks{This work was supported by "SSF Large Intelligent Surfaces - Architecture and Hardware" Project CHI19-0001. Authors are with the Department of Electrical and Information Technology (EIT), Lund University, Sweden (Email: \{ashkan.sheikhi,ove.edfors, juan.vidal\_alegria\}@eit.lth.se).}
}
\maketitle

\begin{abstract}
Non-linear transceivers and non-reciprocity of downlink and uplink channels are two major challenges in the deployment of massive multiple-input-multiple-output (MIMO) systems. We consider an over-the-air (OTA) approach for digital pre-distortion (DPD) and reciprocity calibration to jointly address these issues. In particular, we consider a memory-less non-linearity model for the base station (BS) transmitters, and we propose a method to perform both linearization and reciprocity calibration based on mutual coupling OTA measurements between BS antennas. We show that, by using only the OTA-based data, we can linearize the transmitters and design the calibration to compensate for both the non-linearity and non-reciprocity of BS transceivers. This allows alleviating the requirement to have dedicated hardware modules for transceiver linearization. Moreover, the proposed reciprocity calibration method is solely based on closed-form linear transformations, achieving a significant complexity reduction over state-of-the-art reciprocity methods, which assume linear transceivers, and rely on iterative methods. Simulation results showcase the potential of our approach in terms of the calibration matrix estimation error and downlink data-rates when applying zero-forcing (ZF) precoding after using our OTA-based DPD and reciprocity calibration method.
\end{abstract}


\section{Introduction}
% Introduce BS, UEs, UL, DL, TDD, ZF, LMMSE, MRC, BBU
\IEEEPARstart{M}{assive} \ac{MIMO} has been one of the main technologies in the development of the fifth-generation (5G) of wireless networks, by enabling significant improvements in network capacity and reliability \cite{MassiveMag}. In the early stages of massive \ac{MIMO} development, several proposals motivated the adoption of \ac{FDD} in massive MIMO deployments. While some advantages may arise from considering \ac{FDD} \cite{FDD1}, the over-head in downlink channel estimation is an important drawback which limits the system scalability. Therefore, \ac{TDD} is selected as the more viable approach for the deployment of massive \ac{MIMO} in 5G and beyond, since it enables downlink channel estimation based on uplink \ac{CSI} and channel reciprocity \cite{OveTDDFDD}.

In ideal \ac{TDD} systems, perfect channel reciprocity allows the \ac{BS} to use the \ac{UL} \ac{CSI} for \ac{DL} precoding. However, in practical deployments, the differences between \ac{TX} and \ac{RX} hardware may compromise this assumption \cite{ota_rec_cal}. To compensate the differences, reciprocity calibration methods are employed. There are several approaches for reciprocity calibration in massive \ac{MIMO}. \Ac{OTA}-based reciprocity calibration methods relying on mutual-coupling measurements are specially promising since they do not require dedicated hardware for calibration \cite{joao_recip,ota_rec_cal,nuutti2024acc,NewCal}. Another challenge in implementing massive \ac{MIMO} systems is the non-linear response of the transceivers. There are several methods to compensate these non-linear effects, with per-antenna \ac{DPD} being the most favorable option because of its effectiveness. To perform the \ac{DPD}, many approaches rely on input-output measurements of the \ac{TX}-chains with the aim of designing an inverse function for canceling the non-linear effects \cite{DPD4}. \ac{OTA}-based \ac{DPD} approaches, such as methods based on wireless links with near-field or far-field probes, have emerged as an efficient alternative to linearize the amplifiers \cite{nuutti2024arx}. To the best of our knowledge, no prior work has addressed the joint problem of reciprocity calibration and \ac{TX}-chain linearization using \ac{OTA} techniques. Specifically, existing literature does not explore approaches that unify \ac{DPD} and reciprocity calibration within a shared framework, thereby reducing the need for dedicated \ac{DPD} modules.

In this paper, we propose a method exploiting \ac{OTA} measurements of inter-antenna mutual couplings at the \ac{BS} to perform both the \ac{DPD} and reciprocity calibration. The literature on reciprocity calibration has focused on the latter by assuming linearized transceivers \cite{joao_recip,ota_rec_cal,nuutti2024acc,NewCal}. This assumption is not accurate in practical cases, especially when scaling up massive \ac{MIMO} systems, which necessitates deploying less expensive non-linear components and non-ideal linearization techniques for cost-efficiency reasons. Therefore, we assume that the \ac{TX}-chains in the \ac{BS} are non-linear and we propose a method to linearize them based on mutual coupling measurements. Our approach relies on the same hardware available for \ac{OTA}-based reciprocity calibration, which improves resource efficiency by eliminating the need for dedicated hardware to perform per-antenna \ac{DPD}. Considering the linearized transmitters after applying our \ac{OTA}-based method, we derive a reciprocity calibration approach which depends only on linear closed-form transformations. Thus, the proposed reciprocity calibration allows for a significant complexity reduction over iterative methods focusing on the reciprocity calibration problem \cite{joao_recip,ota_rec_cal,nuutti2024acc,NewCal}. Numerical results show that, even with the extra challenge of having to deal with non-linear \ac{TX}-chain compensation, the calibration matrix estimation error approaches the \ac{CRLB}, which was derived in \cite{joao_recip}. We also evaluate the system performance in downlink data transmission using \ac{ZF} precoding and show that the proposed method can approach the perfect calibration performance even though we are considering non-linear \ac{TX}-chains.

\section{System Model}
% to shorten this, we 
We consider {a \ac{TDD} multi-user-\ac{MIMO} scenario} where an $M$-antenna \ac{BS} serves $K\leq M$ \acp{UE} through a narrow-band channel. For the \acp{UE}, we assume that the \ac{TX}- and \ac{RX}-chains are both operating in the linear regime. For the \ac{BS}, we assume that the \ac{RX}-chains are also operated in the linear regime, but the \ac{TX}-chains exhibit non-linear response.\footnote{Assuming non-linear behavior only in the \ac{BS} \ac{TX}-chains is reasonable taking into account that this is where the input power is significantly higher, pushing the power amplifiers to the non-linear regime \cite{downlink_linear}. For the \acp{UE}, this non-linearity may be compensated with a single DPD module per \ac{UE}.} 

\subsection{Uplink} 
The $M\times 1$ vector of received symbols at the \ac{BS} during an \ac{UL} transmission may be expressed as
\begin{equation}
    \bs{y}_\mrm{B} = \bs{H}_\mrm{UL}\bs{s}_\mrm{U}+\bs{n}_\mrm{B},
\end{equation}
where $\bs{s}_\mrm{U}$ is the $K\times 1$ vector of input symbols to each \ac{UE} \ac{TX}-chain and $\bs{n}_\mrm{B}\sim \mathcal{CN}(\bs{0},N_{0,\mrm{B}}\bs{I}_M)$ models the \ac{AWGN} at the \ac{BS}. The $M\times K$ channel matrix, is given by
\begin{equation}\label{eq:UL_ch}
    \bs{H}_\mrm{UL} = \bs{R}_\mrm{B}\bs{H}\bs{T}_\mrm{U},
\end{equation}
where $\bs{R}_\mrm{B}=\diag(r^{\mrm{B}}_1,\dots,r^{\mrm{B}}_M)$ and $\bs{T}_\mrm{U}=\diag (t_1^\mrm{U},\dots,t^\mrm{U}_K)$ are associated with the linear response of the \ac{BS} receivers and the \ac{UE} transmitters, respectively, and $\bs{H}$ corresponds to the $M\times K$ reciprocal propagation channel matrix \cite{joao_recip}. The \ac{UL} channel in \eqref{eq:UL_ch}, which includes the effects of the \ac{UE} transmitters and the \ac{BS} receivers, can be estimated at the \ac{BS} based on \ac{UL} pilots transmitted by the \acp{UE}, allowing for effective implementation of linear processing techniques, e.g., \ac{ZF} and \ac{MRC}.

Note that, if the UEs transmitters had non-linear behavior, the term $\bs{T}_\mrm{U}$ in the estimated \ac{UL} channel would be substituted by a non-linear function of the pilot matrix. A thorough study of this case may be considered in future work, but the presented method would still be able to cope with the non-linearity and non-reciprocity originated at the \ac{BS} side.

\subsection{Downlink}
During the \ac{DL} transmission phase, the $K\times 1$ vector of symbols received at the \acp{UE} may be expressed as
\begin{equation}\label{eq:MIMO_DL}
    \bs{y}_\mrm{U} = \bs{R}_\mrm{U}\bs{H}^\mrm{T}\bs{f}(\bs{x}_\mrm{B})+\bs{n}_\mrm{U},
\end{equation}
where $\bs{x}_\mrm{B}$ is the $M\times 1$ vector of input symbols to each \ac{BS} \ac{TX}-chain, $\bs{n}_\mrm{U}\sim \mathcal{CN}(\bs{0},\diag(N_{0,\mrm{U}_1},\dots,N_{0,\mrm{U}_K}))$ models the \ac{AWGN} at the \acp{UE}, $\bs{R}_\mrm{U}=\diag((r^{\mrm{U}}_1,\dots,r^{\mrm{U}}_K)$ is associated with the linear response of the \ac{UE} receivers, and $\bs{f}:\mathbb{C}^{M\times 1}  \rightarrow \mathbb{C}^{M \times 1}$ is a vector-valued function modeling the non-linear response of the \ac{BS} \ac{TX}-chains. We assume that the transmitted symbols are generated such that
\begin{equation}\label{eq:tx_symb}
    \bs{x}_\mrm{B} = \bs{g}(\bs{W} \bs{s}_\mrm{B}),
\end{equation}
where $\bs{s}_\mrm{B}$ is the $K\times 1$ vector of symbols intended for the \acp{UE}, $\bs{W}$ is the $M \times K$ linear precoding matrix applied at the \ac{BS} \ac{BBU}, and $\bs{g}:\mathbb{C}^{M\times 1}  \rightarrow \mathbb{C}^{M \times 1}$ is the non-linear vector-valued function associated to the \ac{DPD} applied at each \ac{TX}-chain.

Let us assume that the cross-talk between \ac{TX}-chains is negligible so that $\bs{f}(\cdot)$, and correspondingly $\bs{g}(\cdot)$, are component-wise functions. Considering a third-order memory-less polynomial model \cite{EmilCorrNeg}, we have
\begin{equation}\label{eq:3rd_order}
    f_m(\bs{x}) = t^\mrm{B}_m x_m +\beta_m x_m \vert x_m \vert^2 , \;\;\;\;\forall m \in \{1,\dots,M\},
\end{equation}
where $t^\mrm{B}_m$ and $\beta_m$ are two complex scalars characterizing the non-linear response of the $m$'th \ac{TX}-chain at the \ac{BS}. In general, the \ac{BS} \ac{TX}-chain and \acp{UE} \ac{RX}-chain responses are unknown, which means that the non-linear parameters $t^\mrm{B}_m$ and $\beta_m$, as well as the diagonal entries of $\bs{R}_\mrm{U}$, are unknown at the \ac{BS}. Note that, unlike state-of-the-art work on reciprocity calibration \cite{joao_recip,ota_rec_cal,nuutti2024acc,NewCal}, where $\bs{f}(\cdot)$ is associated to a linear transformation $\bs{T}_{B}=\diag(t^{\mrm{B}}_1,\dots,t^{\mrm{B}}_M)$, we cannot hereby define an aggregated \ac{DL} channel matrix due to the non-linear nature of the TX-chains.

\subsection{Background: OTA Reciprocity Calibration}
As mentioned earlier, previous work has addressed the problem of reciprocity calibration in massive \ac{MIMO} assuming \ac{BS} \ac{TX}-chains operating in linear regime \cite{joao_recip,ota_rec_cal,nuutti2024acc,NewCal}. Under such assumptions, we may define the \ac{DL} channel matrix as
\begin{equation}\label{eq:DL_ch}
    \bs{H}_\mrm{DL} = \bs{R}_\mrm{U}\bs{H}^\mrm{T}\bs{T}_\mrm{B},
\end{equation}
which may be also derived from the presented system model, assuming $\beta_m=0$ in \eqref{eq:3rd_order}.\footnote{Equivalently, reciprocity calibration problems in \cite{joao_recip,ota_rec_cal,nuutti2024acc,NewCal} may be obtained by assuming perfect \ac{DPD} up to unknown scalars, i.e., $\bs{f}(\bs{g}(\bs{x}))=\bs{T}_\mrm{B}\bs{x}$.} The main goal of reciprocity calibration methods is to estimate the reciprocity matrix,
\begin{equation}\label{eq:rec_mat}
    \bs{C} = \bs{T}_\mrm{B}\bs{R}_\mrm{B}^{-1}.
\end{equation}
The reason is that, if we have knowledge of $\bs{C}$, we can transform the estimated UL channel matrix into
\begin{equation}
\begin{aligned}
    \widetilde{\bs{H}}_\mrm{DL} &= (\bs{C}\bs{H}_\mrm{UL})^\mrm{T} \\
    &= \bs{T}_\mrm{U}\bs{H}^\mrm{T} \bs{T}_\mrm{B}.
\end{aligned}
\end{equation}
Note that $\widetilde{\bs{H}}_\mrm{DL}$ corresponds to $\bs{H}_\mrm{DL}$ up to an unknown $K\times K$ diagonal matrix, namely $\bs{D}=\bs{T}_\mrm{U}\bs{R}_\mrm{U}^{-1}$, multiplied from the left. Hence, $\widetilde{\bs{H}}_\mrm{DL}$ can be effectively used for linear precoding, with the only caveat that the symbols received by the \acp{UE} would end up multiplied by an unknown scalar, which has negligible impact on system performance \cite{perf_UE_scalar}.\footnote{In practice, this issue is addressed by sending a DL pilot \cite{joao_recip}.} We may thus ignore the non-reciprocity associated to the \acp{UE} hardware, modeled by $\bs{R}_\mrm{U}$ and $\bs{T}_\mrm{U}$, and focus on characterizing the non-reciprocity associated to the \ac{BS}. An important advantage of the \ac{OTA}-based calibration methods which are based on mutual coupling measurements is that they avoid the need for dedicated hardware to characterize the linear response of each \ac{TX}-chain, and can improve the cost-efficiency of massive \ac{MIMO} systems \cite{joao_recip}. Similarly, we can argue that having dedicated hardware to perform \ac{DPD} may compromise the cost-efficiency of \ac{MIMO} systems with increasing number of antennas, e.g., massive \ac{MIMO} and beyond. Thus, we next propose a method to jointly characterize the non-linear response of the \ac{BS} \ac{TX}-chains, as well as the resulting reciprocity matrix, to suitably design $\bs{g}(\cdot)$ and $\bs{W}$ for effectively serving the \acp{UE} in the \ac{DL}.

\section{OTA DPD and reciprocity calibration} \label{sec:otadpdcal}
Our proposed method may be divided into three stages:
\begin{itemize}
    \item First, the non-linear response of the \ac{BS} \ac{TX}-chains, associated to $\bs{f}(\cdot)$, is estimated based on \ac{OTA} mutual coupling measurements.
    \item Second, the DPD, associated to  $\bs{g}(\cdot)$, is designed based on the estimated non-linear response.
    \item Third, reciprocity calibration is performed based on the DPD-linearized \ac{BS} \ac{TX}-chains, after which effective \ac{DL} precoding, associated to  $\bs{W}$, becomes available at the \ac{BS}.
\end{itemize}
\subsection{OTA non-linearity characterization}
In this stage each \ac{BS} antenna transmits $N_\text{dpd}\geq2$ inter-antenna pilot signals to estimate the non-linearity parameters. The signal received at the $j$'th antenna when the $\ell$'th pilot, $\ell\in \{1,\dots, N_\text{dpd}\}$, is transmitted by the $i$'th antenna may be expressed as
\begin{equation} \label{dpdOTA}
	y_{ij,\ell} = h_{ij}r_j\left(t_ix_{i,\ell} + \beta_ix_{i,\ell}|x_{i,\ell}|^2\right) + n_{ij,\ell} ,
\end{equation}
where $h_{ij}$ is the mutual coupling gain between antennas $i$ and $j$, which is assumed fixed and known at the \ac{BS},\footnote{The coupling gains may be characterized with a single measurement of the antenna system using a network analyzer \cite{joao_recip}. Thus, knowledge of these may be assumed in any MIMO-related scenario with co-located TX antennas.} $x_{i,\ell}$ is the $\ell$'th pilot symbol transmitted by the $i$'th antenna, and $n_{ij,\ell} \sim \mathcal{CN}(0,N_0)$ models the measurement noise. Note that we have removed the superscript B from the parameters $r_j$ and $t_i$ for notation convenience since, as previously reasoned, we may focus on the non-reciprocity associated to the \ac{BS}.

For each pair of non-linearity parameters associated to one \ac{TX}-chain, there are $M-1$ relevant \ac{DPD} measurements per pilot transmission, i.e., all of those originated in the same antenna, but received at different antennas. Thus, each of these measurements would share the same $t_i$ and $\beta_i$ in \eqref{dpdOTA}, but they would be related to a different complex gain $r_j$, associated to the linear response of the \ac{RX}-chain from the respective receiving antenna. Since the complex gains $r_j$ are unknown, it is not possible to directly estimate the non-linearity parameters $t_i$ and $\beta_i$ from this dataset. However, we may combine the $M-1$ measurements by averaging them after compensating for the known mutual coupling gains, so as to reduce the uncertainty, as well as the resulting noise. The combined measurements are then given by
 \begin{equation}\label{dataTild}
 \begin{aligned}
	\Tilde{y}_{i,\ell} &= \frac{1}{M-1}\sum_{j\neq i} \frac{y_{ij,\ell}}{h_{ij}}\\
    &={q_i} (t_i x_{i,\ell}+\beta_i x_{i,\ell} \vert x_{i,\ell}\vert^2)+\tilde{n}_{i,\ell},
\end{aligned}     
 \end{equation}
where the uncertainty is now captured in the unknown parameter $q_i$, given by
\begin{equation}\label{qFactor}
	q_i = \frac{1}{M-1}\sum_{j\neq i} r_j.
\end{equation}
Note that one could also explore alternative optimized combinations to the simple average in \eqref{dataTild}. For example, a weighted average could be optimized assuming a specific model for $h_{ij}$ or a concrete probability distribution for $r_j$, but this is out of scope for this paper and may be considered in future work. {On the other hand, explicit knowledge of $h_{ij}$ could be avoided by absorbing it into $q_i$, as further remarked in Sec. \ref{recipSec}}.

The $N_\text{dpd} \times 1$ data vector $\Tilde{\bs{y}}_{i}=[\Tilde{y}_{i,1},\dots,\Tilde{y}_{i,N_\text{dpd}}]^\mrm{T}$ may then be used to estimate the non-linearity parameters of each antenna up to the unknown factor $q_i$. Since our initial aim is to compensate the non-linear response of the \ac{TX}-chains, this is still possible if we know the non-linear response up to an unknown linear factor, which would only have a linear effect after the non-linearity compensation. In this case, the \ac{DPD} would be designed as if the non-linearity parameters are $\theta_{1i} = q_it_i$ and $\theta_{2i} = q_i\beta_i$. We may thus rewrite the combined data vector as
\begin{equation} \label{LSdpd}
	\bm{\Tilde{y}}_{i} = \bm{\Phi}_i \bm{\theta}_i + \Tilde{\bm{n}}_{i},
\end{equation}
where $\bm{\theta}_i=[\theta_{1i},\theta_{2i}]^\mrm{T}$ is the $2\times 1$ vector of parameters to be estimated, $\bm{\Phi}_i$ is the $N_\text{dpd} \times 2$ known pilot matrix whose columns are given by $\bm{\Phi}_{i,1}=[x_{i,1},\dots,x_{i,N_\text{dpd}}]^\mrm{T}$ and $\bm{\Phi}_{i,2}=[x_{i,1}\vert x_{i,1}\vert^2,\dots,x_{i,N_\text{dpd}}\vert x_{i,N_\text{dpd}}\vert^2]^\mrm{T}$, and $\Tilde{\bm{n}}_{i}\sim \mathcal{CN}(\bs{0},\varsigma_i\bs{I}_{N_\text{dpd}})$ is the resulting noise vector where
\begin{equation} \label{eqNoise}
    \varsigma_i=\frac{N_{0}}{(M-1)^2}\sum_{j\neq i}\frac{1}{ |{h_{ij}}|^2}.
\end{equation}
Since the noise vector is white i.i.d Gaussian, the \ac{LS} estimator is also the \ac{MVU} estimator \cite{est_th}, and can be used to estimate the scaled non-linearity parameters as
\begin{align} \label{estimateDPD}
    \hat{\bs{\theta}}_{i} = ( \bm{\Phi}_i^H \bm{\Phi}_i)^{-1} \bm{\Phi}_i^H \bm{\Tilde{y}}_{i}.
\end{align}
Note that, while we have presented our method for a 3rd order non-linearity model \eqref{eq:3rd_order}, which is the main source of inter-modulation terms falling within the operating frequencies, the method can be generalized for higher order non-linearity models as well. In case of considering non-linearity polynomial models of higher order, the vector $\bm{\theta}$ (correspondingly $\bm{\Phi}$) would include one term per polynomial coefficient and the presented method would still be applicable. Alternatively, one can fit any {RF} non-linear behavior to the 3rd order model, which should still capture its main impact \cite{Ericsson2016}.

\subsection{DPD linearization}
In this stage the non-linearity parameters estimated in the previous stage are used to linearize the output via \ac{DPD}, i.e., by adjusting $\bs{g}(\cdot)$ in \eqref{eq:tx_symb}. The true non-linearity to compensate is the nonlinear function given in \eqref{eq:3rd_order}. However, the estimated non-linearity parameters in \eqref{estimateDPD}, $\theta_{1i}$ and $\theta_{2i}$, characterize a different component-wise function given by
\begin{equation}
\begin{aligned}
    \tilde{f}_m(\bs{x}) &= \theta_{1m}x_m+\theta_{2m}x_m|x_m|^2 \\
    &= q_m f_m(\bs{x}).
\end{aligned}
\end{equation}
We may thus express
\begin{equation}\label{eq:true_NL}
    \bs{f}(\bs{x}) = \bs{Q}^{-1} \tilde{\bs{f}}(\bs{x}),
\end{equation}
where $\bs{Q}=\diag(q_1,\dots,q_M)$.

Since the function $\tilde{\bs{f}}(\cdot)$ is fully characterized, we can find its inverse by using methods such as the postdistortion approach \cite{EunInverse}. 
We may then select
\begin{equation}\label{eq:inverse}
    \bs{g}(\cdot)=\tilde{\bs{f}}^{-1}(\cdot),
\end{equation}
which is applied to the the transmitted symbols as described in \eqref{eq:tx_symb}. In practice, perfect \ac{DPD} inversion may not be fully achievable, mainly due to limited \ac{DPD} size and imperfect estimation of non-linearity parameters. We have considered imperfect inversion in the numerical results from Section~\ref{sec_OTAWCL_num}.

The resulting symbols transmitted through the reciprocal channel, may then be expressed as
\begin{equation}
\begin{aligned}
    \bs{f}(\bs{x}_\mrm{B}) &= \bs{Q}^{-1} \tilde{\bs{f}}(\bs{g}(\bs{W}\bs{s}_\mrm{B})) \\
    &= \bs{Q}^{-1} \bs{W}\bs{s}_\mrm{B}.
\end{aligned}
\end{equation}
Hence, applying the proposed \ac{OTA}-\ac{DPD}, results in an equivalent linear transmitter gain given by $\Tilde{\bs{T}}_\mrm{B}=\bs{Q}^{-1}$. Now that the transmitter is linear, we can define a \ac{DL} channel matrix equivalent to \eqref{eq:DL_ch}, but substituting $\bs{T}_\mrm{B}$ for $\Tilde{\bs{T}}_\mrm{B}$, so that reciprocity calibration methods as those presented in \cite{joao_recip,ota_rec_cal} are directly applicable. However, we will show that the reciprocity calibration can be performed without the need for complex iterative methods.

\subsection{Reciprocity calibration}\label{recipSec}
The last stage consists of performing \ac{OTA}-based calibration considering the \ac{TX}-chains previously linearized through the \ac{OTA}-\ac{DPD} stages. To this end, each \ac{BS} antenna transmits pilots to other antennas. The received symbols at the $j$'th antenna from the $i$'th antenna may be expressed as
\begin{equation}\label{eq:OTA_cal}
    {y}_{ij} = h_{ij}r_j\Tilde{t}_i {x}_{ij} + n_{ij},
\end{equation}
where the variables have direct correspondence with those defined in \eqref{dpdOTA}, but substituting $t_i$ for {$\tilde{t}_i = 1/{q_i}$} and having $\beta_i=0$. The measurements defined in \eqref{eq:OTA_cal}
can be directly employed to estimate the product of unknown parameters $r_j\Tilde{t}_i$. In fact, we may now use the trivial \ac{MVU} estimator, given by
\begin{equation}\label{eq:est_rtt}
    \widehat{r_j\Tilde{t}_i} = \frac{1}{h_{ij}{x}_{ij}} {y}_{ij}.
\end{equation}
However, in order to perform reciprocity calibration, we are actually interested in the reciprocity parameters, $c_m={\tilde{t}_m}/{r_m}$, which define the adjusted reciprocity matrix entries from \eqref{eq:rec_mat}.

In \cite{joao_recip}, it was noted that multiplying all the reciprocity parameters by a common scalar does not compromise the effectiveness of the reciprocity calibration.\footnote{This constant scalar may be absorbed in the linear response of the \acp{UE} \ac{RX}-chains, given by $\bs{R}_\mrm{U}$ in \eqref{eq:MIMO_DL}.} Thus, we may select one of the calibration parameters, e.g., $c_1$, and normalize all the rest by that value. The resulting scaled calibration parameters may then be expressed as
\begin{equation}\label{cal_tild}
\begin{aligned} 
    \Tilde{c}_m &\triangleq \frac{c_m}{c_1}=\frac{r_1 \Tilde{t}_m}{r_m \Tilde{t}_1}
\end{aligned}
\end{equation}
which corresponds to the ratio of $r_j\Tilde{t}_i$ products appearing in \eqref{eq:est_rtt} for $(i,j)\in\{(1,m),(m,1)\}$. Since each of these products can be estimated through \eqref{eq:est_rtt}, we can find estimates for the scaled calibration parameters by
\begin{equation}\label{eq:ctildhat}
    \widehat{\Tilde{c}_m} = \frac{\widehat{r_1 \Tilde{t}_m}}{\widehat{r_m \Tilde{t}_1}}.
\end{equation}
The estimation error of $\widehat{\Tilde{c}_i}$ can be reduced by averaging several estimates of $r_1 \Tilde{t}_i$ and $r_i \Tilde{t}_1$, which is possible if each \ac{BS} antenna transmits $N_\text{cal}\geq2$ pilots in \eqref{eq:OTA_cal}. Note further that, assuming reciprocity of the mutual coupling coefficients, i.e. $h_{ij}=h_{ji}$, we could still estimate $\Tilde{c}_m$ using \eqref{eq:OTA_cal} without explicit knowledge of $h_{ij}$ since the coefficients would cancel each other in \eqref{eq:ctildhat}. The method does not rely on this assumption, since the coupling coefficients can be estimated in practice. Nevertheless, the coupling coefficients between antennas are reciprocal per definition since we are absorbing the non-reciprocity within our model.

We have thus shown that we can estimate all the entries of the calibration matrix up to a constant, i.e., we can estimate $\Tilde{\bs{C}} = \frac{1}{c_1} \bs{C}$, by means of simple linear estimators. This allows achieving reciprocity without the need for high complexity iterative algorithms, such as the algorithms used in \cite{joao_recip}.
\section{Numerical results}\label{sec_OTAWCL_num}
In this section, we perform simulations to validate the feasibility and assess the performance of the proposed method. The number of \ac{BS} antennas and the number of single-antenna \acp{UE} are $M=100$ and $K=10$, respectively. For the \ac{BS} TX-chains non-linearity parameters in \eqref{eq:3rd_order}, we fit a 3rd order polynomial to the measurement data from \cite{Ericsson2016} for a Gallium Nitride (GaN) amplifier operating at 2.1 GHz at a sample rate of 200 MHz and a signal bandwidth of 40 MHz. For the RX-chains complex gains, we use the values in \cite{joao_recip} given by $r_m^B = 0.9+0.2\frac{M-m}{M}\exp(j2\pi m/M)$. To implement the imperfect inverse function in the \ac{DPD}, we generate a look-up table based on the \ac{OTA} data. For the mutual coupling channel gains in \eqref{dpdOTA}, we have used the linear \ac{LS} fit based on the measurements in \cite{joao_recip}. We also define the \ac{OTA} link \ac{SNR} as the receive \ac{SNR} for the link between the antennas with least mutual coupling gain.

\begin{figure}[t]
	\centering	\includegraphics[width=1\linewidth]{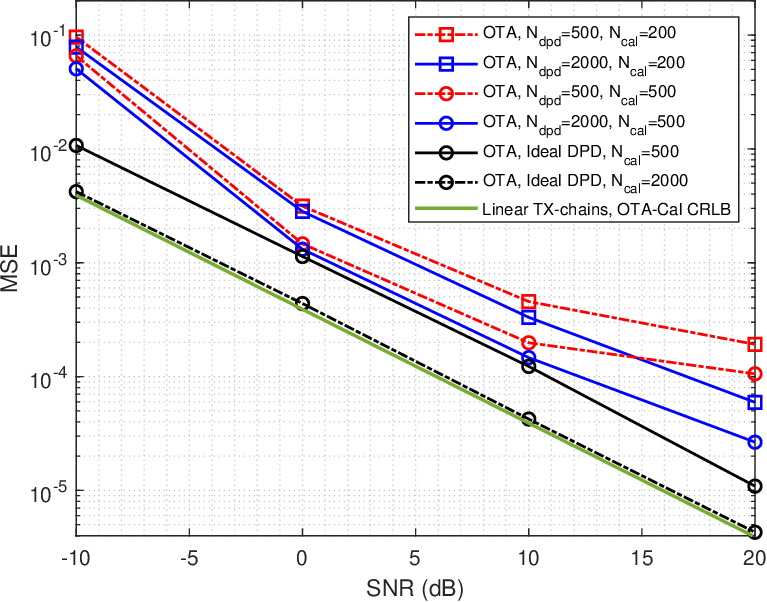}
	\caption{Average \ac{MSE} of calibration matrix estimation using the proposed \ac{OTA} \ac{DPD} and reciprocity calibration.}
	\label{MSE}
\end{figure}

Fig.~\ref{MSE} illustrates the average \ac{MSE} of the calibration matrix estimation with our proposed \ac{OTA}-based method for different levels of \ac{SNR}. The calibration matrix is estimated after performing the \ac{OTA}-\ac{DPD}, with $N_\text{dpd}=500$ or $2000$ \ac{OTA} transmissions. In the calibration step, we have considered $N_\text{cal}=200$, $500$, or $2000$ \ac{OTA} transmissions. Note that transceiver characteristics are slowly-varying parameters, which means that even larger values of $N_\text{dpd}$ and $N_\text{cal}$ would still have a rather small impact on the total overhead. For comparison we have included an upper bound for the performance of the reciprocity calibration problem at hand, \ac{CRLB}, derived in \cite{joao_recip}, which assumes linear \ac{TX}-chains. We have also included the case with an ideal \ac{DPD} followed by the \ac{OTA}-based reciprocity calibration. Firstly, we can see that the performance of the \ac{OTA}-based \ac{DPD} and reciprocity calibration approaches the ideal \ac{DPD} case. Secondly, we can see that the performance is fairly close to the \ac{CRLB}, even though we have the extra challenge of dealing with non-linear \ac{TX}-chains. Note that, in order to approach the \ac{CRLB}, which is the ultimate performance bound for the reciprocity problem, \cite{joao_recip} requires an iterative algorithm of considerable complexity. The reported complexity order in \cite{joao_recip}, which is considered state-of-the-art in massive MIMO reciprocity calibration \cite{State_Cal}, is given by $\mathcal{O}(M^2N_\text{ite})$, where $N_\text{ite}$ is the number of iterations of the algorithm. The complexity of our reciprocity calibration method is given by $\mathcal{O}(M N_\text{cal})$, since it requires averaging $N_\text{cal}$ numbers where each of them is obtained by performing 2 multiplications for each of the $M-1$ antennas. Thus, for massive \ac{MIMO} and beyond, where $M\gg1$, our method may even attain significant complexity reduction, since reducing $N_\text{cal}$ only has a minor impact on the calibration matrix MSE, as seen from Fig.~\ref{MSE}. Other state-of-the-art reciprocity calibration methods require even higher complexity, while their performance is still limited by the \ac{CRLB}. For example, the one in \cite{NewCal} for mm-Wave systems requires $\mathcal{O}(M^3)$ complexity to tackle the non-reciprocity problem with linear transceivers.

\begin{figure}[t]
	\centering
	\includegraphics[width=0.97\linewidth]{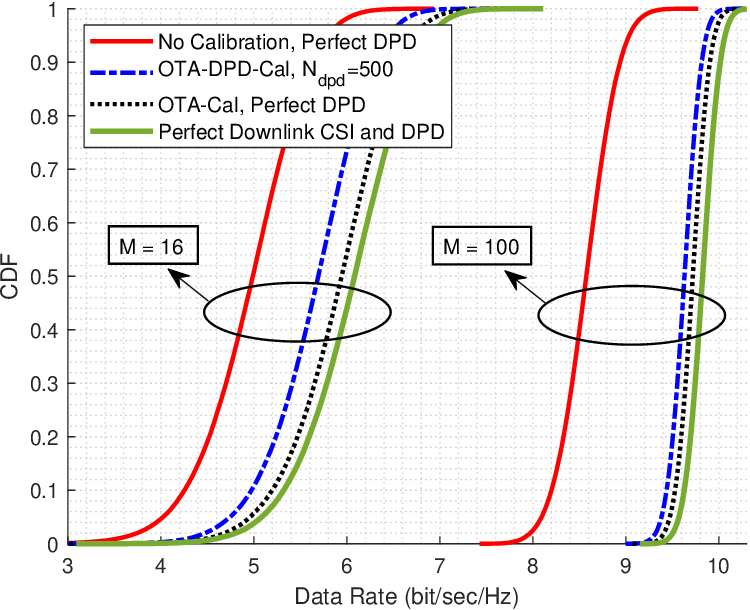}
	\caption{CDF of a \ac{UE} \ac{DL} data rate.}
	\label{Rate_CDF}
\end{figure}

Fig.~\ref{Rate_CDF} illustrates the CDF of \ac{DL} data rate under \ac{ZF} precoding, where we have generated $10^4$ \ac{DL} realizations of an i.i.d. Rayleigh fading channel, and the \ac{DL} signal power for each \ac{UE} is selected to achieve an average \ac{SNR} of $10$ dB at their receivers. We have selected $N_\text{cal}=500$ for the \ac{OTA} reciprocity calibration, \ac{OTA} reference \ac{SNR} of $0$ dB, and \ac{OTA}-\ac{DPD} with $N_\text{dpd}=500$. We have also performed the same reciprocity calibration for a case with perfect \ac{DPD}. For comparison, we have included two extreme cases both with a perfect \ac{DPD}, one with perfect \ac{DL} \ac{CSI} (ideal calibration), and one worst-case scenario with no calibration. We can observe that the limited-size \ac{DPD} performs very close to the perfect \ac{DPD} case, which further confirms the observations from  Fig.~\ref{MSE}. As for the calibration performance, we can see that the proposed \ac{OTA}-\ac{DPD} and reciprocity calibration approaches the ideal case with perfect \ac{DL} \ac{CSI}, without requiring high-complexity iterative algorithms, even-though we are dealing with the extra challenge of non-linear \ac{TX}-chains. The gain from adopting the proposed \ac{OTA}-case is more significant for higher number of antennas, i.e., massive \ac{MIMO} and beyond, and the \ac{OTA}-\ac{DPD} performs closer to the ideal \ac{DPD} case.

\section{Conclusion}
In this paper, we have proposed an \ac{OTA}-based method for \ac{DPD} and reciprocity calibration in massive \ac{MIMO} and beyond. In particular, we considered a memory-less non-linearity model for the \ac{BS} transmitters and proposed to perform the linearization and reciprocity calibration by using \ac{OTA} measurements of the mutual coupling among the \ac{BS} antennas. We showed that, by only using the \ac{OTA} data, we can effectively linearize the transmitters and perform reciprocity calibration with reduced complexity over state-of-the-art. Simulation results showed promising performance of the proposed methodology, both in terms of the calibration matrix estimation error and the \ac{DL} data-rates when applying \ac{ZF} precoding after our \ac{OTA}-based \ac{DPD} and calibration method.

\bibliographystyle{IEEEtran}
\bibliography{IEEEabrv,Sheikhi_WCL2025-1687_Refs}

\end{document}